\documentclass[a4, conference]{IEEEtran}
\usepackage{cite}
\usepackage{pgfplots}

\pgfplotsset{compat=1.10}

\usepgfplotslibrary{fillbetween}

\usepackage{bm}
\usepackage{lipsum}
\usepackage{makeidx}
\usepackage{enumerate}
\usepackage{color}
\usepackage{cite}
\usepackage{amsmath,amsthm}   
\usepackage{amssymb}
\usepackage{nomencl}
\usepackage{multirow}
\usepackage{graphicx}
\usepackage{epstopdf}
\usepackage{threeparttable}
\usepackage{multicol}
\DeclareGraphicsExtensions{.pdf,.jpeg,.png,.jpg,.emf,.eps}
\usepackage{calc}
\usepackage{here}
\usepackage{url}

\usepackage{upref}
\usepackage{comment}
\usepackage{times}
\usepackage{dsfont}
\usepackage{epic,eepic}
\usepackage{rawfonts}
\usepackage[T1]{fontenc}
\usepackage{latexsym}
\usepackage{amsfonts}

\usepackage{capitalgreekitalic}

\usepackage[font=small,labelfont=bf]{caption}
\usepackage[font=small,labelfont=bf]{subcaption}

\usepackage{xcolor}
\usepackage{tikz,pgfplots}
\usetikzlibrary{calc}
\usetikzlibrary{intersections}
\usetikzlibrary{arrows,shapes}
\usetikzlibrary{positioning}

\theoremstyle{definition}

\allowdisplaybreaks

\begin{document}
\title{
Energy-efficient Rate Splitting for MIMO STAR-RIS-assisted  Broadcast  Channels \\ with I/Q Imbalance
}
\vspace{-.7cm}
\author{%
   \IEEEauthorblockN{Mohammad Soleymani$^*$, Ignacio Santamaria$^\dag$, and Eduard Jorswieck$^\ddag$}
   \IEEEauthorblockA{*Signal \& System Theory Group, Universit\"at Paderborn, Germany \\
                     $^\dag$Dept. Communications Engineering, Universidad de Cantabria, Spain\\
$^\ddag$ Institute for Communications Technology, Technische Universit\"at Braunschweig, Germany\\
                     Email: \small{\protect\url{mohammad.soleymani}@sst.upb.de}, \small{\protect\url{i.santamaria@unican.es}}, \small{\protect\url{jorswieck@ifn.ing.tu-bs.de}}
}
} \vspace{-.9cm}
\maketitle
\begin{abstract}
This paper proposes an energy-efficient scheme for multicell multiple-input, multiple-output (MIMO) simultaneous transmit and reflect (STAR) reconfigurable intelligent surfaces (RIS)-assisted broadcast channels by employing rate splitting (RS) and improper Gaussian signaling (IGS). Regular RISs can only reflect signals. Thus, a regular RIS can assist only when the transmitter and receiver are in the reflection space of the RIS. However, a STAR-RIS can simultaneously transmit and reflect, thus providing a $360^\circ$ coverage.  In this paper, we assume that transceivers may suffer from I/Q imbalance (IQI). To compensate for IQI, we employ IGS. Moreover, we employ RS to manage intracell interference. We show that RIS can significantly improve the energy efficiency (EE) of the system when RIS components are carefully optimized.  Additionally, we show that STAR-RIS can significantly outperform a regular RIS when the regular RIS cannot cover all the users.   We also show that RS can highly increase the EE comparing to treating interference as noise.
\end{abstract} 
\begin{IEEEkeywords}
 Energy efficiency,  improper Gaussian signaling, majorization minimization, MIMO broadcast channels, rate splitting, reflecting intelligent surface.
\end{IEEEkeywords}

\section{Introduction}
Among the main targets of 6G is to improve the energy efficiency (EE) by a factor of $10\sim 100$ times higher than the EE of 5G networks \cite{gong2022holographic}. A promising technology to improve EE is reconfigurable intelligent surface (RIS), which can provide extra degrees of freedom by optimizing the propagation channels or generally the environment \cite{wu2021intelligent, di2020smart}. Another promising technology to improve EE and/or spectral efficiency (SE) is rate splitting  (RS), which is a powerful interference-management technique that includes many other techniques such as treating interference as noise (TIN), spatial division multiple access (SDMA), and non-orthogonal multiple access (NOMA) \cite{mao2022rate}. This paper employs RS and RIS to enhance the EE of a multi-cell broadcast channel (BC).

RIS has been shown to improve EE and SE of various networks \cite{huang2019reconfigurable, wu2019intelligent, pan2020multicell, soleymani2022improper, weinberger2022synergistic, jiang2022interference, santamaria2023icassp}.  
RIS can modulate channels, thus improving the coverage or reducing/neutralizing the interference. Hence, there can be different applications for RIS in a network. For instance, RIS can manage interference in some specific scenarios such as $K$-user interference channels \cite{jiang2022interference, santamaria2023icassp}. Moreover, RIS can strengthen the cell-edge links (either in interference-free or interference-limited systems), which can highly improve the minimum rate and or EE of the network.   

 A regular passive RIS can only reflect signals, which may restrict its coverage range. Indeed, both transceivers should be in the reflection space of a regular RIS so that it can modulate the channel. To address this issue, simultaneously reflect and transmit RIS (STAR-RIS) is proposed that can reflect and transmit simultaneously, which provides a full $360^\circ$ coverage \cite{9774942, liu2022simultaneously, mu2021simultaneously, xu2021star, soleymani2023spectral}. Thus, STAR-RIS may be able to support a wider range of applications compared to a regular RIS. 

RIS can manage intercell interference in a multi-cell BC. However, RIS cannot fully mitigate intracell interference especially when a network is over loaded, i.e., when the number of users per cell is higher than the number of base station (BS) antennas \cite{soleymani2022improper, soleymani2022rate, soleymani2022noma}.  Thus, the system performance can be improved if other interference management techniques such as RS are employed in RIS-assisted systems. We refer the reader to \cite{mao2022rate} for a more detailed overview on RS. 

Another main obstacle to improve EE and/or SE is hardware impairment (HWI). In practice, devices are never completely ideal, which can significantly degrade the system performance especially when such impairments are not compensated \cite{soleymani2020rate, soleymani2020improper, javed2019multiple}. A source of HWI is an imbalance in in-phase and quadrature branches, which is known as I/Q imbalance (IQI) \cite{javed2019multiple, soleymani2020improper}. When IQI occurs, the output signal can be modeled as a widely linear transformation of the input signals, which may make the signals and noise improper. To compensate for IQI, we can employ improper Gaussian signaling (IGS), which is also an interference-management technique \cite{cadambe2010interference, soleymani2020improper,  soleymani2019energy,  
soleymani2019improper, soleymani2023rate, 
javed2020journey}.

In this paper, we employ the 1-layer RS alongside with IGS to improve EE of a multicell STAR-RIS-assisted BC with IQI. In \cite{soleymani2022rate}, we proposed a framework for RS in RIS-assisted systems by mainly studying a regular RIS. In this work, we focus on STAR-RIS and propose energy-efficient RS schemes for different operational modes of STAR-RIS by considering three different feasibility sets. We consider two schemes for STAR-RIS. First, we assume that each STAR-RIS component can simultaneously transmit and reflect, which is known as energy splitting (ES). Second, we randomly divide STAR-RIS components into two sets. In the first set, we assume that all components can only reflect, while in the second set, all components can only transmit. This scheme is known as mode switching (MS). We show that the joint RS and RIS design significantly increases the EE of the system. Moreover, we show that STAR-RIS can highly outperform a regular RIS with a relatively low number of RIS components when the regular RIS cannot cover all the users. The MS scheme can perform very close to the ES scheme for more realistic feasibility sets for RIS components in the considered scenario. 
\section{System model}\label{sec-ii}
We consider a multi-cell MIMO STAR-RIS-assisted BC with $L$ multiple-antenna BSs, each serving $K$ users with $N_u$ receive antennas. Each BS has $N_{BS}$ transmit antennas. 
We assume that there are $M$ passive STAR-RISs with $N_{RIS}$ elements to assist the communications.
We further assume that each transceiver suffers from IQI based on the model in \cite{javed2019multiple}. For notational simplicity, we consider a symmetric scenario without a loss of generality. However, the proposed scheme can be easily extended to asymmetric systems.  

\subsection{STAR-RIS model}
\begin{figure}
    \centering
\includegraphics[width=.4\textwidth]{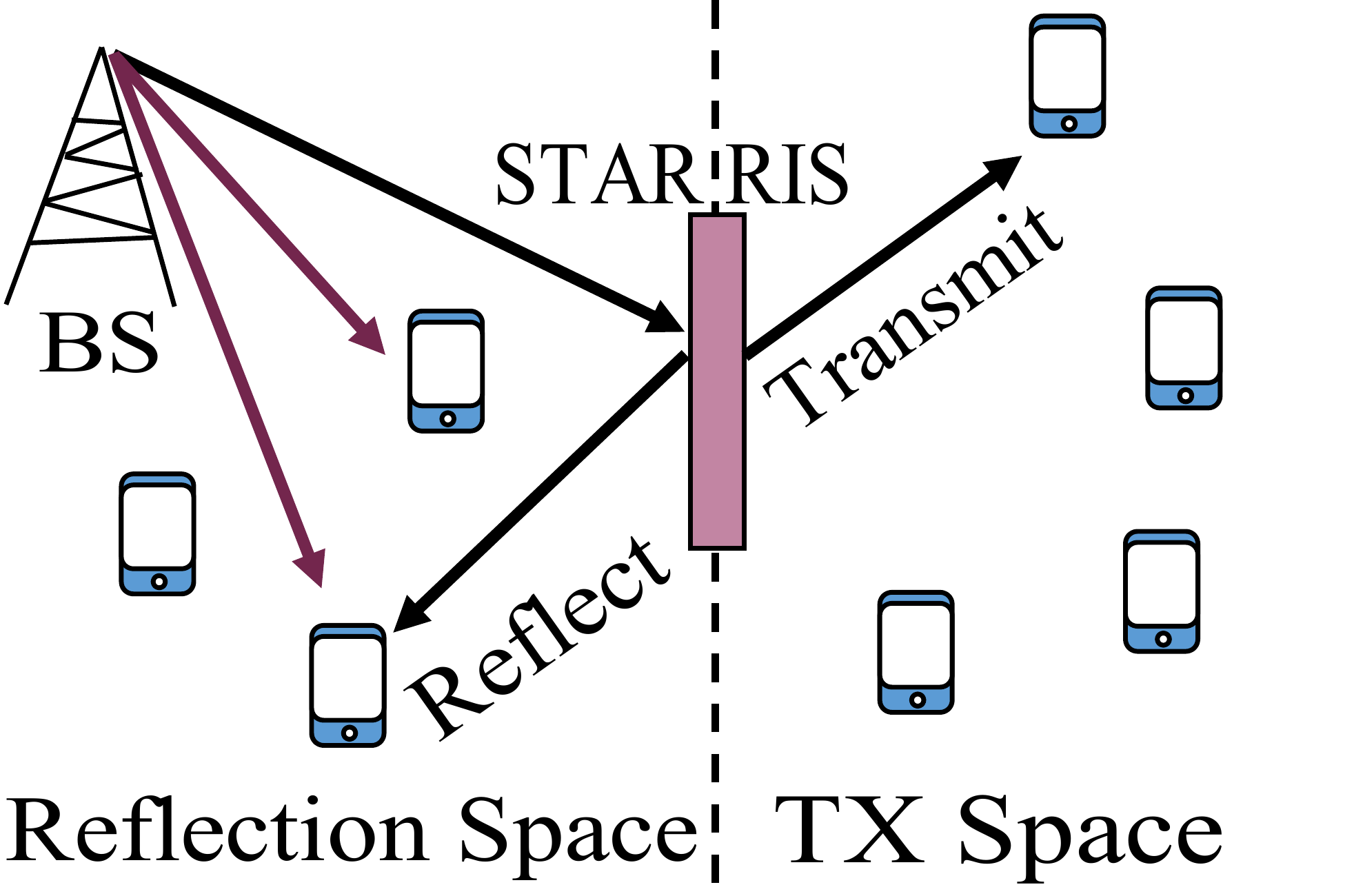}
    \caption{A typical STAR-RIS-assisted system.}
	\label{Fig-st}  
\end{figure}
In a STAR-RIS-assisted system, each user belongs to either reflection space or transmission space  (see Fig. \ref{Fig-st}). Hence, the channel between BS $i$ and the $k$-th user associated to BS $l$ (denoted by u$_{lk}$) is  \cite[Eq. (2)]{mu2021simultaneously} 
\begin{equation}
\mathbf{H}_{lk,i}
=\underbrace{\sum_{m=1}^M\mathbf{G}_{lk,m}\bm{\Theta}_m^{r/t}\mathbf{G}_{mi}}_{\text{links through STAR-RIS}}+
\underbrace{\mathbf{F}_{lk,i}}_{\text{direct link}},
\end{equation}
where $\mathbf{G}_{lk,m}$ is the channel matrix between the $m$-th RIS and u$_{lk}$,  $\mathbf{F}_{lk,i}$ is the direct channel between BS $i$ and u$_{lk}$, and $\mathbf{G}_{mi}$ is the channel matrix between RIS $m$ and BS $i$. 
The matrix $\bm{\Theta}_m^r$ (or $\bm{\Theta}_m^t$) is a diagonal matrix containing the reflection (or transmission) coefficient of each RIS component. In other words,
$\bm{\Theta}_m^r
=\text{diag}\left(\theta_{m_1}^r, \theta_{m_2}^r,\cdots,\theta_{m_{N_{RIS}}}^r\right)$, and 
$\bm{\Theta}_m^t
=\text{diag}\left(\theta_{m_1}^t, \theta_{m_2}^t,\cdots,\theta_{m_{N_{RIS}}}^t\right)$, 
where $\theta_{m_i}^r$ (or $\theta_{m_i}^t$) is the reflection (or transmission) coefficient of the $i$-th STAR-RIS component. If a user is in the reflection (or transmission) space of a STAR-RIS, the channel for the user can be controlled by  $\theta_{m_i}^r$ (or $\theta_{m_i}^t$).
Unfortunately, the reflection and transmission coefficients of each RIS component  are dependent. 
There are three models for the relation between $\theta_{m_i}^r$ and $\theta_{m_i}^t$. In the first model, the amplitudes of $\theta_{m_i}^r$ and $\theta_{m_i}^t$ are related to each other as in the set \cite[Eq. (2)]{xu2021star}
\begin{equation}\label{model-2}
\mathcal{T}_{U}=\left\{\theta_{m_i}^r,\theta_{m_i}^t:|\theta_{m_i}^{r}|^2+|\theta_{m_i}^{t}|^2\leq 1 \,\,\,\forall m,i\right\}.
\end{equation}
The set $\mathcal{T}_{U}$ is convex. In another model,  the amplitudes of $\theta_{m_i}^r$ and $\theta_{m_i}^t$ are related to each other as in the  set \cite[Eq. (1)]{liu2021star}
\begin{equation}\label{model-1}
\mathcal{T}_{I}=\left\{\theta_{m_i}^r,\theta_{m_i}^t:|\theta_{m_i}^{r}|^2+|\theta_{m_i}^{t}|^2= 1 \,\,\,\forall m,i\right\}.
\end{equation}
In the sets $\mathcal{T}_{U}$ and $\mathcal{T}_{I}$, the phases of $\theta_{m_i}^r$ and $\theta_{m_i}^t$ can be independently optimized.  In another model, not only the amplitudes of $\theta_{m_i}^r$ and $\theta_{m_i}^t$ are dependent, but also their phases are related as in the following set \cite{9774942}, \cite[Lemma 1]{soleymani2023spectral}
\begin{multline*}
\mathcal{T}_{N}\!\!=\!\!\left\{\!\theta_{m_i}^r,\theta_{m_i}^t\!\!:\!|\theta_{m_i}^{r}|^2\!+\!|\theta_{m_i}^{t}|^2\!=\! 1,
|\theta_{m_i}^{r}\!\pm\!\theta_{m_i}^{t}|^2\!\leq\! 1,
\forall m,i\right\}\!\!.
\end{multline*}

There are three different modes for operating a STAR-RIS. First, each STAR-RIS component can simultaneously reflect and transmit, which is known as the ES scheme. Second, a set of STAR-RIS components operate only in the transmission mode and the other components operate only the reflection mode, which is known as the MS scheme. Third, all STAR-RIS components operate in the transmission mode in a time slot, and then they operate in the reflection mode in the next time slot, which is known as the time switching (TS) scheme. In this paper, we do not consider the TS scheme since it operates similar to a regular RIS in each time slot, and may not provide a considerable gain over a regular RIS when time sharing among users is not allowed \cite{soleymani2023spectral}. 

\subsection{Signal model}
In this paper, we consider the 1-layer RS in which each BS transmits two types of messages, i.e., common and private. Thus, the BS intends to transmit 
$\mathbf{x}_{l}=\mathbf{x}_{c,l}+\sum_{k}\mathbf{x}_{lk}$, 
where $\mathbf{x}_{c,l}$ is the common message, which is decoded by all the users associated to the $l$-th BS. Moreover, $\mathbf{x}_{lk}$ is the private message for u$_{lk}$. The signals $\mathbf{x}_{c,l}$ and $\mathbf{x}_{lk}$ for all $l,k$ are zero-mean independent improper Gaussian signals.
As BS $l$ suffers from IQI, the actual transmitted signal of BS $l$ is a widely linear transformation of  $\mathbf{x}_{l}$, i.e., $\mathbf{x}_{tl}=\bm{\Gamma}_{1t}\mathbf{x}_{l}+\bm{\Gamma}_{2t}\mathbf{x}_{l}^*$, where $\bm{\Gamma}_{1t}$ and $\bm{\Gamma}_{2t}$ are given by \cite[Eq. (7)]{soleymani2020improper} and \cite[Eq. (8)]{soleymani2020improper}, respectively. The received signal at u$_{lk}$ is 
${\mathbf{y}}_{rlk}=\sum_{i=1}^L{\mathbf{H}}_{lk,i}
\left(\bm{\Gamma}_{1t}\mathbf{x}_{l}+\bm{\Gamma}_{2t}\mathbf{x}_{l}^*\right)+{\mathbf{n}}_{lk}$, 
where ${\mathbf{n}}_{lk}$ is zero-mean white additive proper Gaussian noise. Due to IQI at the receiver of u$_{lk}$, the actual received signal at u$_{lk}$  is a widely linear transform of  ${\mathbf{y}}_{rlk}$ as $\mathbf{y}_{lk}=\bm{\Gamma}_{1r}{\mathbf{y}}_{rlk}+\bm{\Gamma}_{2r}{\mathbf{y}}_{rlk}^{*}$, where $\bm{\Gamma}_{1r}$ and $\bm{\Gamma}_{2r}$ are given by \cite[Eq. (12)]{soleymani2020improper} and \cite[Eq. (13)]{soleymani2020improper}, respectively. To deal with impropriety, we employ the real-decomposition method in which every signal/parameter is written in a real domain.   
Employing \cite[Lemma 1]{soleymani2020improper}, the received signal at u$_{lk}$ can be written as
$\underline{\mathbf{y}}_{lk}
=
{ \sum_{i=1
}^L\underline{\mathbf{H}}_{lk,i}
\underline{\mathbf{x}}_{i}}
+
{\underline{\mathbf{n}}_{lk}}$, where the parameters are defined as in \cite[Eq. (2)]{soleymani2022rate}.

\subsection{Rate and energy efficiency expressions}
The rate of u$_{lk}$ is the summation of the achievable rate of its private message, $\mathbf{x}_{lk}$, and its dedicated rate from the common message, $\mathbf{x}_{c,l}$, as
$r_{lk}=r_{p,lk}+r_{c,lk}$,
where $r_{p,lk}$ is the rate of 
 the private message $\mathbf{x}_{lk}$ after decoding and canceling the common message $\mathbf{x}_{c,l}$ while treating the other interfering signals as noise, i.e.,
\begin{align*}
r_{p,lk}\!=\!\frac{1}{2}\!\log_2\!\left|{\bf I}\!+\!
{\bf D}_{lk}^{-1}{\bf S}_{lk}
 \right|\!\!=\!
\underbrace{
\frac{1}{2}\log_2\!\left|{\bf D}_{lk}\!\!
+\!{\bf S}_{lk}
\right|
}_{r_{p,lk_1}
}\!
-\!
\underbrace{
\frac{1}{2}\!\log_2\left|{\bf D}_{lk}
\right|}
_{r_{p,lk_2}
},
\end{align*}
where ${\bf S}_{lk}=\underline{{\bf H}}_{lk,l}{\bf P}_{lk}\underline{{\bf H}}_{lk,l}^T $, 
and 
\begin{align*}
\mathbf{D}_{lk}
&=
\underbrace{
\sum_{i=1,i \neq l}^L\underline{\mathbf{H}}_{lk,i}
\mathbf{P}_{i}\underline{\mathbf{H}}_{lk,i}^T
}_{\text{Intercell interference}}
+
\underbrace{
\sum_{j= 1,j\neq k}^{K}
\underline{\mathbf{H}}_{lk,l}
\mathbf{P}_{lj}
\underline{\mathbf{H}}_{lk,l}^T
}_{\text{Intracell interference}}
+
\underbrace{
\underline{\mathbf{C}}_{n}}_{\text{Noise}},
\end{align*}
where $\underline{\mathbf{C}}_{n}$ is the covariance matrix of the noise $\underline{\mathbf{n}}_{lk}$, i.e., $\underline{\mathbf{C}}_{n}=\mathbb{E}\left\{\underline{\mathbf{n}}_{lk}\underline{\mathbf{n}}_{lk}^T\right\}$, and $\mathbf{P}_i=\mathbf{P}_{c,i}+\sum_k\mathbf{P}_{ik}$ for all $i$, where $\mathbf{P}_{c,i}$, and ${\bf P}_{ik}$ are, respectively, the transmit covariance matrix of  $\underline{{\bf x}}_{c,i}$ and $\underline{{\bf x}}_{ik}$.
  The term $r_{c,lk}\geq 0$ in 
$r_{lk}$ refers to the portion of the common rate allocated to u$_{lk}$. The common message $\underline{{\bf x}}_{c,l}$ should be decodable by all users associated to BS $l$. Thus, the transmission rate of the common message at BS $l$ should be  chosen according to\vspace{-.1cm}
\begin{equation}\label{fis=}
r_l=\sum_{k=1}^{K}r_{c,lk}\leq \min_{
k}\left\{\bar{r}_{c,lk}\right\}\triangleq r_{c,l},
\end{equation}
 where $\bar{r}_{c,lk}$ is the maximum decodable rate of $\underline{{\bf x}}_{c,l}$ at u$_{lk}$ treating the other signals as noise 
\begin{align*}
\bar{r}_{c,lk}
=
\underbrace{
\frac{1}{2}\log_2\left|
{\bf D}_{c,lk}
+
{\bf S}_{c,lk}
\right|
}_{\bar{r}_{c,lk_1}
}
-
\underbrace{
\frac{1}{2}\!\log_2\left|
{\bf D}_{c,lk}
\right|}
_{\bar{r}_{c,lk_2}
},
\end{align*}
where ${\bf S}_{c,lk}=
\underline{{\bf H}}_{lk,l}
{\bf P}_{l,c}\underline{{\bf H}}_{lk,l}^T$ and ${\bf D}_{c,lk}={\bf S}_{lk}+{\bf D}_{lk}$.

The EE of a user is defined as the ratio between its rate and total consumed energy to transmit its data as  \cite{zappone2015energy}
\begin{equation}
e_{lk}=\frac{r_{lk}}{P_c+\eta \text{Tr}\left(\mathbf{P}_{lk}+\mathbf{P}_{c,l}/K\right)},
\end{equation}
where $1/\eta$ is the power efficiency of the BSs and $P_c$ is the constant power consumption for transmitting
data to a user, given by \cite[Eq. (27)]{soleymani2022improper}.
\subsection{Problem statement}
In this paper, we maximize the minimum-weighted EE (MWEE) of users:\vspace{-.1cm}
\begin{subequations}\label{opt}
\begin{align}
\!\underset{\{\mathbf{P}\}\in\mathcal{P},\{\bm{\Theta}\}\in\mathcal{T},\mathbf{r}_c,e
 }{\max}\!\! & 
  e\!\!\!\!\!\!
&\text{s.t.}\,\,&e_{lk}\geq \alpha_{lk}e,\,\,\forall lk,
\\
&&&
\!\sum_k\!r_{c,lk}\!\leq\! \min_{
k}\left\{\bar{r}_{c,lk}\right\}\!\!=\!r_{c,l},\!\forall l,
\\
&&&
r_{c,lk}\geq 0,\,\,r_{lk}\geq r_{lk}^{th},\,\,\forall lk,
\end{align}
\end{subequations}
where $\mathbf{r}_c=\{r_{c,lk}\}_{\forall lk}$, $\{\mathbf{P}\}=\{\mathbf{P}_{lk},\mathbf{P}_{c,l}\}_{\forall lk}$, and $\{\bm{\Theta}\}=\{\bm{\Theta}_{m}^r,\bm{\Theta}_{m}^t\}_{\forall m}$ are the optimization variables. 
Moreover, $\alpha_{lk}$ is the weight for u$_{lk}$, which corresponds to the priority of the user. 
The set $\mathcal{P}$ includes all the feasible transmit covariance matrices and is given by \cite[Eq. (3)]{soleymani2022rate}. Moreover, $\mathcal{T}$ is the feasibility set for RIS components, which can be  $\mathcal{T}_U$, $\mathcal{T}_I$, or $\mathcal{T}_N$.
\section{Proposed rate splitting scheme}\label{sec-iii}
\vspace{-.1cm}
The optimization problem \eqref{opt} is non-convex and difficult. In this section, we find a suboptimal solution for \eqref{opt} by employing an iterative approach based on majorization minimization (MM) and alternating optimization (AO) to approximate \eqref{opt}. 
That is, we first fix the RIS components, $\{\bm{\Theta}^{(t-1)}\}=\{\bm{\Theta}^{r^{(t-1)}}\!\!,\bm{\Theta}^{t^{(t-1)}}\}$, and solve \eqref{opt} over transmit covariance matrices to obtain $\{\mathbf{P}^{(t)}\}$. 
We then solve \eqref{opt} over RIS components for given transmit covariance matrices, $\{\mathbf{P}^{(t)}\}$, to find $\{\bm{\Theta}^{(t)}\}$.
We iterate this procedure until a convergence condition is met.

\subsection{Optimizing transmit covariance matrices}
The optimization problem \eqref{opt} is non-convex even for fixed $\{\bm{\Theta}^{(t-1)}\}$ since the rates are not concave in $\{\mathbf{P}\}$. 
To solve \eqref{opt} for fixed $\{\bm{\Theta}^{(t-1)}\}$, we first find  suitable surrogate functions for the rates by employing convex-concave procedure (CCP) since the rates can be written as a difference two concave functions in $\{\mathbf{P}\}$. To this end, we employ \cite[Lemma 3]{soleymani2022rate}, which results in
\begin{align*}
 r_{p,lk}\geq& \tilde{r}_{p,lk}=
r_{p,lk_1}\left(\{\mathbf{P}\}\right) 
-r_{p,lk_2}^{(t-1)}
\\
&-\sum_{j=1,\neq k}^{K}\!\!\!
\text{Tr}\!\left(
\frac{
\underline{\mathbf{H}}_{lk,l}^T
(\mathbf{D}_{lk}^{(t-1)})^{-1}
\underline{\mathbf{H}}_{lk,l}
}
{2\ln 2}
\left(\mathbf{P}_{lj}-\mathbf{P}_{lj}^{(t-1)}\right)
\right)
\\
&-\sum_{i=1, \neq l}^L
\text{Tr}\left(
\frac{
\underline{\mathbf{H}}_{lk,i}^T
(\mathbf{D}_{lk}^{(t-1)})^{-1}
\underline{\mathbf{H}}_{lk,i}
}
{2\ln 2}
\left(\mathbf{P}_{i}-\mathbf{P}_{i}^{(t-1)}\right)
\right),
\\
\bar{r}_{c,lk}\geq& \tilde{r}_{c,lk}=
r_{c,lk_1}\left(\{\mathbf{P}\}\right) 
-r_{c,lk_2}^{(t-1)}
\\
&-\sum_{j=1}^{K}
\text{Tr}\left(
\frac{
\underline{\mathbf{H}}_{lk,l}^T
(
\mathbf{D}_{c,lk}^{(t-1)})^{-1}
\underline{\mathbf{H}}_{lk,l}
}
{2\ln 2}
\left(\mathbf{P}_{lj}-\mathbf{P}_{lj}^{(t-1)}\right)
\right)
\\
&-\sum_{i=1, \neq l}^L
\text{Tr}\left(
\frac{
\underline{\mathbf{H}}_{lk,i}^T
(
\mathbf{D}_{c,lk}^{(t-1)})^{-1}
\underline{\mathbf{H}}_{lk,i}
}
{2\ln 2}
\left(\mathbf{P}_{i}-\mathbf{P}_{i}^{(t-1)}\right)
\right),
\end{align*}
where $r_{p,lk_2}^{(t-1)}=r_{p,lk_2}\left(\{\mathbf{P}^{(t-1)}\}
\right)$, $\mathbf{D}_{lk}^{(t-1)}=\mathbf{D}_{lk}\left(\{\mathbf{P}^{(t-1)}\}
\right)$, $r_{c,lk_2}^{(t-1)}=r_{c,lk_2}\left(\{\mathbf{P}^{(t-1)}\}
\right)$, and $\mathbf{D}_{c,lk}^{(t-1)}=\mathbf{D}_{c,lk}\left(\{\mathbf{P}^{(t-1)}\}
\right)$. Substituting the rates with the concave lower bounds in \eqref{opt} for fixed $\{\bm{\Theta}^{(t-1)}\}$ yields
\begin{subequations}\label{opt-p2}
\begin{align}
\!\underset{\{\mathbf{P}\}\in\mathcal{P},\mathbf{r}_c,e 
 }{\max}\!\! & 
  e\!\!\!\!\!\!
&\text{s.t.}\,\,&
\frac{\tilde{r}_{lk}}{P_c\!+\!\eta \text{Tr}\left(\mathbf{P}_{lk}\!+\mathbf{P}_{c,l}/K\right)}\geq \alpha_{lk}e,\,\,\forall lk,
\\
&&&
\!\sum_k\!r_{c,lk}\!\leq\! \min_{
k}\left\{\tilde{r}_{c,lk}\right\}\!\!=\!r_{c,l},\!\forall l,
\\
&&&
r_{c,lk}\geq 0,\,\,\tilde{r}_{lk}\geq r_{lk}^{th},\,\,\forall lk,
\end{align}
\end{subequations}
which is non-convex, but its global optimal solution can be obtained by the generalized Dinkelbach algorithm \cite{zappone2015energy}. Due to a space restriction, we do not provide details and refer the reader to, e.g.,  \cite{zappone2015energy}.
\subsection{Optimizing RIS components}
In the following, we first consider the ES scheme. Then, we mention how the scheme can be adapted to the MS scheme, which is a special case of  the ES scheme.
For fixed $\{\mathbf{P}^{(t)}\}$, \eqref{opt} is non-convex since the rates are not concave in $\{\bm{\Theta}\}$, and the set $\mathcal{T}$ is non-convex for $\mathcal{T}_I$ and $\mathcal{T}_N$. To solve \eqref{opt} for fixed $\{\mathbf{P}^{(t)}\}$, we first obtain a concave lower bound for the rates by 
applying \cite[Lemma 2]{soleymani2022improper} as
\begin{align*} 
r_{p,lk}&\!\!\geq\!
 \hat{r}_{p,lk}\!\!=\!r_{p,lk}^{(t-1)}\!\!-
\frac{1}{2\ln 2}\!\left(\text{{Tr}}\!\left(
\bar{\mathbf{S}}_{lk}\bar{\mathbf{D}}_{lk}^{-1}\!
\right)
\!+
2\!\!
\text{{ Tr}}\!\left(\!
\bar{\mathbf{V}}_{lk}^T
\bar{\mathbf{D}}_{lk}^{-1}\mathbf{V}_{lk}\!
\right)\right.
\\
&-
\left.
\text{{Tr}}\!\left(\!
(\bar{\mathbf{D}}^{-1}_{lk}\!-\!(\bar{\mathbf{S}}_{lk}\! +\! \bar{\mathbf{D}}_{lk})^{-1})^T
(\mathbf{S}_{lk}\!+\!\mathbf{D}_{lk})\!\right)
\right)\!,
\\
\bar{r}_{c,lk}&\!\!\geq\!
 \hat{r}_{c,lk}\!\!=\!r_{c,lk}^{(t-1)}\!\!-
\frac{1}{2\ln 2}\!\left(\text{{Tr}}\!\left(
\bar{\mathbf{S}}_{c,lk}\bar{\mathbf{D}}_{c,lk}^{-1}\!
\right)
\!+\!
2
\text{{Tr}}\!\left(\!
\bar{\mathbf{V}}_{c,lk}^T
\bar{\mathbf{D}}_{c,lk}^{-1}\right.\right.
\\&
\left.\left.
\!\!\!\times \mathbf{V}_{c,lk}\!
\right)
\!-\!
\text{{Tr}}\!\left(\!\!
(\bar{\mathbf{D}}^{-1}_{c,lk}\!-\!(\bar{\mathbf{S}}_{c,lk}\! +\! \bar{\mathbf{D}}_{c,lk})^{-1})^T
(\mathbf{S}_{c,lk}\!+\!\mathbf{D}_{c,lk})\!\right)
\!\!\right)\!\!,
\end{align*}
where  $\mathbf{V}_{lk}=\underline{{\bf H}}_{lk,l}\mathbf{P}_{lk}^{(t)^{1/2}}$, $\bar{\mathbf{V}}_{lk}=\underline{{\bf H}}_{lk,l}\left(\{\bm{\Theta}\}^{(t-1)}\right)\mathbf{P}_{lk}^{(t)^{1/2}}$, $\bar{\mathbf{D}}_{lk}=\mathbf{D}_{lk}\left(\{\bm{\Theta}\}^{(t-1)}\right)$, 
$\mathbf{V}_{c,lk}=\underline{{\bf H}}_{lk,l}\mathbf{P}_{c,l}^{(t)^{1/2}}$, $\bar{\mathbf{V}}_{c,lk}=\mathbf{S}_{c,lk}^{{1/2}}\left(\{\bm{\Theta}\}^{(t-1)}\right)$, and 
$\bar{\mathbf{D}}_{c,lk}=\mathbf{D}_{c,lk}\left(\{\bm{\Theta}\}^{(t-1)}\right)$. 
Inserting the concave lower bounds for the rates into \eqref{opt} for fixed $\{\mathbf{P}^{(t)}\}$ results in
\begin{subequations}\label{opt-t2}
\begin{align}
\underset{\{\bm{\Theta}\}\in\mathcal{T},\mathbf{r}_c,e
 }{\max}\!\! & 
  e\!\!\!\!\!\!
&\text{s.t.}\,\,&\frac{\hat{r}_{lk}}{P_c\!+\!\eta \text{Tr}\left(\mathbf{P}_{lk}\!+\!\mathbf{P}_{c,l}/K\right)}\!\geq \alpha_{lk}e,\,\forall lk,
\label{12a}
\\
&&&
\!\sum_k\!r_{c,lk}\!\leq\! \min_{
k}\left\{\hat{r}_{c,lk}\right\}\!\!=\!r_{c,l},\!\forall l,
\label{12b}
\\
&&&
r_{c,lk}\geq 0,\,\,\hat{r}_{lk}\geq r_{lk}^{th},\,\,\forall lk,
\label{12c}
\end{align}
\end{subequations}
which is convex only when $\mathcal{T}$ is a convex set, i.e., when $\mathcal{T}=\mathcal{T}_U$. In this case, the proposed algorithm converges to a stationary point of \eqref{opt} for both ES and MS schemes since the surrogate functions for the rates fulfill the three conditions in \cite[Sec. III]{soleymani2020improper}. We propose a suboptimal approach to convexify $\mathcal{T}_I$ and $\mathcal{T}_N$ below.

The sets $\mathcal{T}_I$ and $\mathcal{T}_N$ are not convex due to the constraint $|\theta_{m_i}^{r}|^2+|\theta_{m_i}^{t}|^2= 1$, which can be written as the two constraints: $|\theta_{m_i}^{r}|^2+|\theta_{m_i}^{t}|^2\leq 1$ and $|\theta_{m_i}^{r}|^2+|\theta_{m_i}^{t}|^2\geq 1$. The former is a convex constraint, but the latter is not since $|\theta_{m_i}^{r}|^2+|\theta_{m_i}^{t}|^2$ is a convex function, instead of being concave. 
Thus, we can apply CCP to convexify this constraint as
\begin{multline}\label{eq-50-6}
\zeta(\theta_{m_i}^r,\theta_{m_i}^t)
\triangleq
|\theta_{m_i}^{r^{(t-1)}}|^2+2\mathfrak{R}\left(\theta_{m_i}^{r^{(t-1)}}(\theta_{m_i}^r-\theta_{m_i}^{r^{(t-1)}})^*\right)+
\\
|\theta_{m_i}^{t^{(t-1)}}|^2
+2\mathfrak{R}\left(\theta_{m_i}^{t^{(t-1)}}(\theta_{m_i}^t-\theta_{m_i}^{t^{(t-1)}})^*\right)\geq 1.
\end{multline}
To make the convergence faster, we can relax \eqref{eq-50-6} by introducing $\epsilon>0$ as $\zeta(\theta_{m_i}^r,\theta_{m_i}^t)\geq 1-\epsilon$.
Thus, we can approximate \eqref{opt-t2} for $\mathcal{T}_I$ as
\begin{subequations}\label{opt-t3}
\begin{align}
\underset{\{\bm{\Theta}\}\in\mathcal{T},\mathbf{r}_c,e
 }{\max}\!\! & 
  e\!\!\!\!\!\!\!\!\!\!\!\!
&\text{s.t.}\,\,&\eqref{12a}-\eqref{12c},\,\zeta(\theta_{m_i}^r,\theta_{m_i}^t)\!\!\geq\!\! 1-\!\epsilon,
\forall m_i,
\\
&&&
|\theta_{m_i}^{r}|^2+|\theta_{m_i}^{t}|^2\leq 1,\,\,\,
\forall m_i,
\end{align}
\end{subequations}
which is convex and can be efficiently solved. Note that for the feasibility set $\mathcal{T}_N$, we should add the convex constraints $|\theta_{m_i}^{r}\pm\theta_{m_i}^{t}|^2\leq 1$ for all $m_i,$ to \eqref{opt-t3} to ensure that $\angle \theta_{m_i}^{r}=\angle \theta_{m_i}^{r}\pm \pi/2$. Since these constraints are convex, the resulting problem is convex for $\mathcal{T}_N$.
We call the solution of \eqref{opt-t3} for $\mathcal{T}_I$ (or $\mathcal{T}_N$)
as $\bm{\Theta}_m^{r^{(\star)}}$ and $\bm{\Theta}_m^{t^{(\star)}}$, which may not be feasible due to the relaxation. 
To obtain a feasible point, we project $\bm{\Theta}_m^{r^{(\star)}}$ and $\bm{\Theta}_m^{t^{(\star)}}$ onto $\mathcal{T}_{I}$ (or $\mathcal{T}_N$) as
\begin{align}\label{eq3300}
\hat{\theta}_{m_i}^t\!&\!=\!\frac
{{\theta}_{m_i}^{t^{(\star)}}}
{\sqrt{|{\theta}_{m_i}^{t^{(\star)}}|^2+|{\theta}_{m_i}^{r^{(\star)}}|^2}},&\!\!
\hat{\theta}_{m_i}^r\!&\!=\!\frac
{{\theta}_{m_i}^{r^{(\star)}}}
{\sqrt{|{\theta}_{m_i}^{t^{(\star)}}|^2+|{\theta}_{m_i}^{r^{(\star)}}|^2}},
\end{align}
for all $m_i$. 
Finally, we update $\bm{\Theta}_m^r$ and $\bm{\Theta}_m^t$ such that the solution of the scheme is non-decreasing, i.e., as 
\begin{equation}\label{eq-42-star}
\{\!\bm{\Theta}^{(t)}\!\}\!=\!\!
\left\{\!\!\!\!\!
\begin{array}{lcl}
\{\hat{\bm{\Theta}}\}\!\!&\!\!\!\!\!\!\text{if}\!\!&
\min_{lk}\left\{e_{lk}\!\!\left(\{\hat{\bm{\Theta}}\}\!\right)/\alpha_{lk}\right\}\!\!\geq\!\!\!\!\!
\\
&&
\min_{lk}\left\{e_{lk}\!\!\left(\{\bm{\Theta}^{(t-1)}\}\!\right)/\alpha_{lk}\right\}
\\
\{\bm{\Theta}^{(t-1)}\}&&\text{Otherwise},
\end{array}
\right.
\end{equation}
where $\{\hat{\bm{\Theta}}\}=\{\hat{\bm{\Theta}}^{r},\hat{\bm{\Theta}}^{t}\}$ is given by \eqref{eq3300}. 
The convergence of the proposed scheme is ensured by \eqref{eq-42-star}.

In the MS scheme, we assume that a set of the RIS components operate only in the reflection mode  ($\theta_{m_i}^t =0$), while the other RIS components operate only in the transmission mode  ($\theta_{m_i}^r = 0$). Hence, the proposed scheme for the ES scheme can be applied to the MS scheme by enforcing ${\theta}_{m_i}^t$ (or ${\theta}_{m_i}^r$) to zero for the set of RIS components, operating in the reflection (or transmission) mode. 

Note that the number of  optimization parameters decreases by half in the MS scheme, which reduces the computational complexities of the scheme. Additionally, from a practical point of view, it might be easier to operate in only transmission or reflection mode for a STAR-RIS component.
However, the MS scheme can be considered as lower bound for the performance of the ES scheme since MS is a special case of ES. In other words, an optimal ES scheme never performs worse than any MS scheme since the solution of an MS scheme is a valid solution for an ES scheme. 
\section{Numerical results}\label{sec-iv}
In this section, the simulation parameters are chosen similar to \cite{soleymani2022improper}. 
\begin{figure}
    \centering
\includegraphics[width=.4\textwidth]{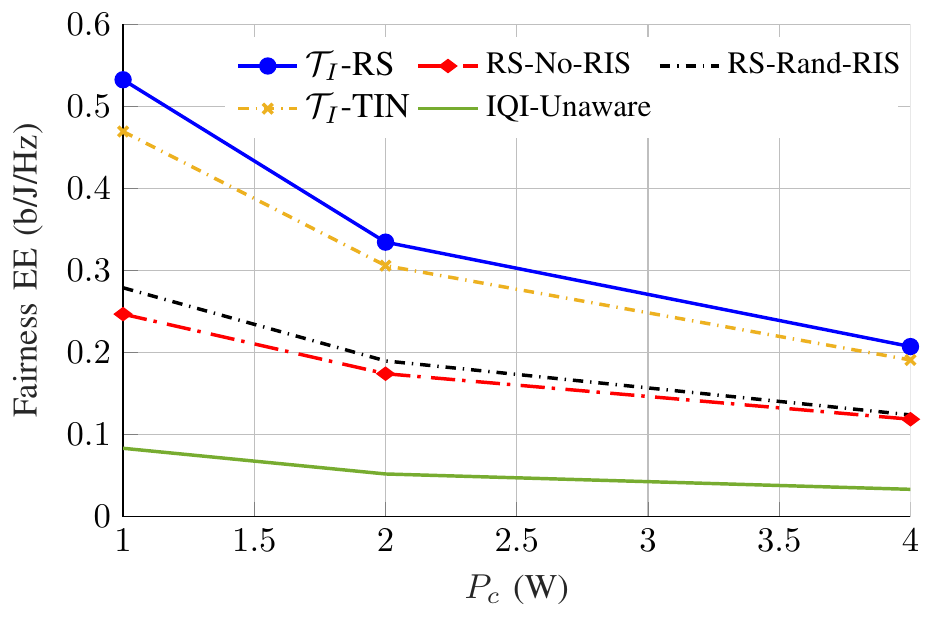}
    \caption{The average fairness EE versus $P_c$ for $N_{BS}=N_u=2$,  $N_{RIS}=20$, $L=2$, $M=2$ and  $K=4$.}
	\label{Fig-2}  
\end{figure}
In Fig. \ref{Fig-2}, we   show the average fairness EE versus $P_c$ for $N_{BS}=N_u=2$,  $N_{RIS}=20$, $L=2$, $M=2$ and $K=2$. In this figure, we consider a two-cell system with one RIS in each cell, as shown in \cite[Fig. 3]{soleymani2022improper}. We assume that all the users are in the reflection space of the RIS to show the impact of RIS and RS on the EE of the users. As can be observed, RIS can significantly improve the EE of the system. Moreover, it can be observed that IQI can highly degrade the system performance if it is not considered in the system design. Note that the IQI-unaware scheme employs proper Gaussian signaling with random RIS components, while the other schemes employ IGS. We also observe that RS can considerably outperform TIN from an EE point of view. 

\begin{figure}
    \centering
\includegraphics[width=.4\textwidth]{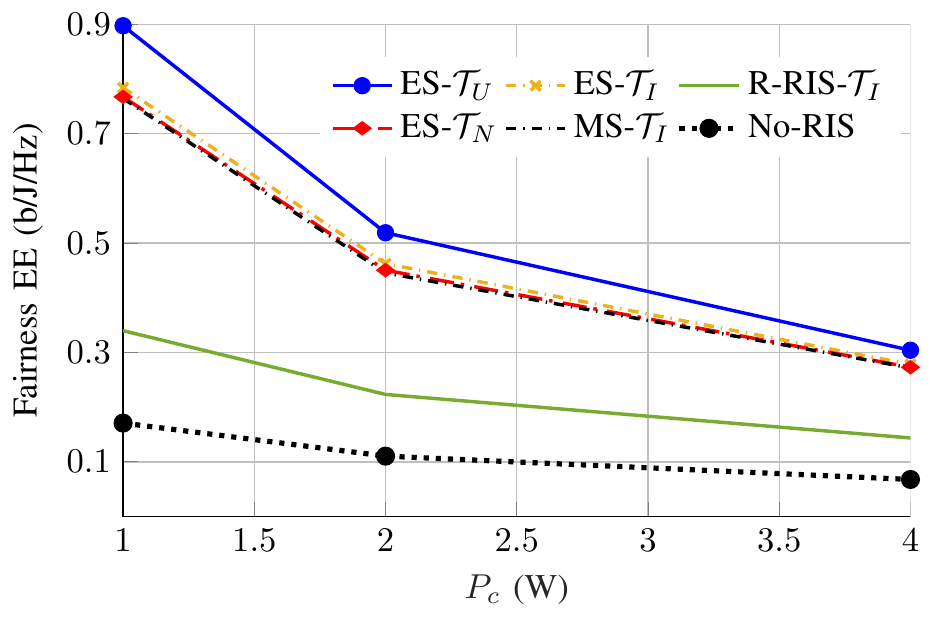}
    \caption{The average fairness EE versus $P_c$ for $N_{BS}=N_u=2$,  $L=1$, $M=1$, $K=8$ and $N_{RIS}=100$.} 
	\label{Fig-3}  
\end{figure}
Fig. \ref{Fig-3} shows the average fairness EE versus $P_c$ for $N_{BS}=N_u=2$,  $L=1$, $M=1$, $K=8$ and $N_{RIS}=100$. In this figure, we compare the performance of a regular RIS with a STAR-RIS in a single-cell BC. To this end, we assume that both regular and STAR-RISs employ the same number of RIS components. We also assume that half of the users are in the reflection space and the other half are in the transmission space of RISs, meaning that the regular RIS can cover only half of users, while the STAR-RIS can assist all the users. As can be observed, RIS can highly enhance EE even when the regular RIS is unable to cover all the users, 
Furthermore, STAR-RIS can significantly outperform a regular RIS. The benefits of STAR-RIS increase with the number of users and the number of RIS components. Additionally, we observe that STAR-RIS with the MS scheme with a unit modulus constraint performs close to the ES scheme with $\mathcal{T}_I$ and $\mathcal{T}_N$.

\section{Conclusion}\label{sec-v}
We proposed energy-efficient schemes for multi-cell STAR-RIS BCs by employing RS and IGS. We considered imperfect devices, suffering from IQI. We showed that RIS can significantly improve the EE of the system. Additionally, we showed that STAR-RIS can highly outperform a regular RIS when the regular RIS is unable to cover all the users. Moreover, we showed that RS can enhance the EE of the system. Finally, our  results show that IQI can highly degrade the EE of the system when it is not compensated in the design.
\section*{Acknowledgment}
The work of I. Santamaria has been partly  supported by the project ADELE PID2019-104958RB-C43, funded by MCIN/AEI/10.13039/501100011033. The work of Eduard Jorswieck was supported in part by the Federal Ministry of Education and Research (BMBF, Germany) as part of the 6G Research and Innovation Cluster 6G-RIC under Grant 16KISK031.
\bibliographystyle{IEEEtran}
\bibliography{ref2}
\end{document}